\documentclass[sn-mathphys,Numbered]{sn-jnl}

\usepackage{graphicx}%
\usepackage{multirow}%
\usepackage{amsmath,amssymb,amsfonts}%
\usepackage{amsthm}%
\usepackage{mathrsfs}%
\usepackage[title]{appendix}%
\usepackage{xcolor}%
\usepackage{textcomp}%
\usepackage{manyfoot}%
\usepackage{booktabs}%
\usepackage{algorithm}%
\usepackage{url}
\usepackage{algorithmicx}%
\usepackage{algpseudocode}%
\usepackage{listings}%

\begin{document}

\title[Article Title]{Segment Anything in Medical Images}


\author[1,2,3]{\fnm{Jun} \sur{Ma}}\email{junma.ma@mail.utoronto.ca}

\author[4]{\fnm{Yuting} \sur{He}}\email{yhe668@uwo.ca}

\author[1]{\fnm{Feifei} \sur{Li}}\email{ff.li@mail.utoronto.ca}

\author[5]{\fnm{Lin} \sur{Han}}\email{me@linhan.email}

\author[6]{\fnm{Chenyu} \sur{You}}\email{chenyu.you@yale.edu}

\author*[1,2,3,7,8]{\fnm{Bo} \sur{Wang}}\email{bowang@vectorinstitute.ai}

\affil[1]{Peter Munk Cardiac Centre, University Health Network, Toronto, Canada}

\affil[2]{Department of Laboratory Medicine and Pathobiology, University of Toronto, Toronto, Canada}

\affil[3]{Vector Institute, Toronto, Canada}

\affil[4]{Department of Computer Science, Western University, Canada}


\affil[5]{Tandon School of Engineering, New York University, USA}

\affil[6]{Department of Electrical Engineering, Yale University, USA}

\affil[7]{Department of Computer Science, University of Toronto, Toronto, Canada}

\affil[8]{UHN AI Hub, Toronto, Canada}


\abstract{Medical image segmentation is a critical component in clinical practice, facilitating accurate diagnosis, treatment planning, and disease monitoring. However, existing methods, often tailored to specific modalities or disease types, lack generalizability across the diverse spectrum of medical image segmentation tasks. Here we present MedSAM, a foundation model designed for bridging this gap by enabling universal medical image segmentation. The model is developed on a large-scale medical image dataset with 1,570,263 image-mask pairs, covering 10 imaging modalities and over 30 cancer types. We conduct a comprehensive evaluation on 86 internal validation tasks and 60 external validation tasks, demonstrating better accuracy and robustness than modality-wise specialist models. By delivering accurate and efficient segmentation across a wide spectrum of tasks, MedSAM holds significant potential to expedite the evolution of diagnostic tools and the personalization of treatment plans.
}




\maketitle

\section*{Introduction}\label{sec1}
Segmentation is a fundamental task in medical imaging analysis, which involves identifying and delineating regions of interest (ROI) in various medical images, such as organs, lesions, and tissues~\cite{nnunet21}. Accurate segmentation is essential for many clinical applications, including disease diagnosis, treatment planning, and monitoring of disease progression~\cite{NatMed-Retina2018,heart-nature}. 
Manual segmentation has long been the gold standard for delineating anatomical structures and pathological regions, but this process is time-consuming, labor-intensive, and often requires a high degree of expertise. Semi- or fully-automatic segmentation methods can significantly reduce the time and labor required, increase consistency, and enable the analysis of large-scale datasets~\cite{wang-deepigeos-PAMI}. 

Deep learning-based models have shown great promise in medical image segmentation due to their ability to learn intricate image features and deliver accurate segmentation results across a diverse range of tasks, from segmenting specific anatomical structures to identifying pathological regions~\cite{msd-paper}. However, a significant limitation of many current medical image segmentation models is their task-specific nature. These models are typically designed and trained for a specific segmentation task, and their performance can degrade significantly when applied to new tasks or different types of imaging data~\cite{seg-reviewPAMI}. 
This lack of generality poses a substantial obstacle to the wider application of these models in clinical practice. In contrast, recent advances in the field of natural image segmentation have witnessed the emergence of segmentation foundation models, such as Segment Anything Model (SAM)~\cite{2023-SAM-Meta} and Segment Everything Everywhere with Multi-modal prompts all at once~\cite{2023-SEEM}, showcasing remarkable versatility and performance across various segmentation tasks. 

There is a growing demand for universal models in medical image segmentation: models that can be trained once and then applied to a wide range of segmentation tasks. Such models would not only exhibit heightened versatility in terms of model capacity, but also potentially lead to more consistent results across different tasks. 
However, the applicability of the segmentation foundation models (e.g., SAM~\cite{2023-SAM-Meta}) to medical image segmentation remains limited due to the significant differences between natural images and medical images. 
Essentially, SAM is a promptable segmentation method that requires points or bounding boxes to specify the segmentation targets. This resembles conventional interactive segmentation methods~\cite{wang-deepigeos-PAMI,wang-interactive2,InteractiveSegMIA, luo2021mideepseg} but SAM has better generalization ability, while existing deep learning-based interactive segmentation methods focus mainly on limited tasks and image modalities. 

Many studies have applied the out-of-the-box SAM models to typical medical image segmentation tasks~\cite{SAM-pathology, SAM-LiverTumor, SAM-Meds, SAM-DKFZ-Abdomen, SAM-Polyps, SAM-BrainMR} and other challenging scenarios~\cite{SAM-Infrared, SAM-Camouflaged, SAM-Camouflaged2, SAM-RealworldApplication}.
For example, the concurrent studies~\cite{SAM-MedIA,SAM-Med-XinYang} conducted a comprehensive assessment of SAM across a diverse array of medical images, underscoring that SAM achieved satisfactory segmentation outcomes primarily on targets characterized by distinct boundaries. However, the model exhibited substantial limitations in segmenting typical medical targets with weak boundaries or low contrast. In congruence with these observations, we further introduce MedSAM, a refined foundation model that significantly enhances the segmentation performance of SAM on medical images. MedSAM accomplishes this by fine-tuning SAM on an unprecedented dataset with more than one million medical image-mask pairs.

We thoroughly evaluate MedSAM through comprehensive experiments on 86 internal validation tasks and 60 external validation tasks, spanning a variety of anatomical structures, pathological conditions, and medical imaging modalities.
Experimental results demonstrate that MedSAM consistently outperforms the state-of-the-art (SOTA) segmentation foundation model~\cite{2023-SAM-Meta}, while achieving performance on par with, or even surpassing specialist models~\cite{nnunet21,deeplabV3plus} that were trained on the images from the same modality. 
These results highlight the potential of MedSAM as a new paradigm for versatile medical image segmentation. 

\begin{figure}[htbp]
\centering
\includegraphics[width=\linewidth]{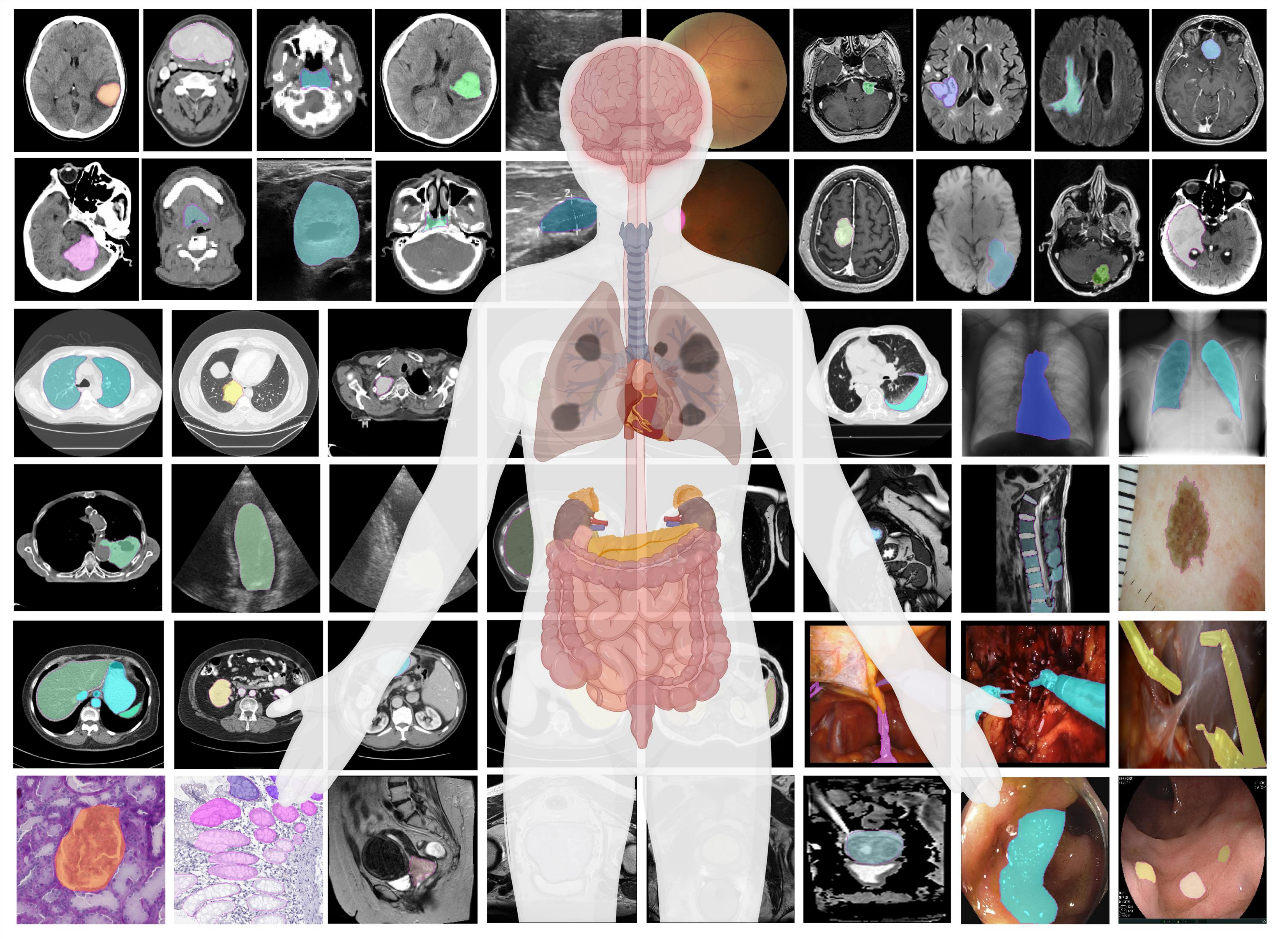}
\caption{\textbf{MedSAM is trained on a large-scale dataset that can handle diverse segmentation tasks.} The dataset covers a variety of anatomical structures, pathological conditions, and medical imaging modalities. The magenta contours and mask overlays denote the expert annotations and MedSAM segmentation results, respectively.}
\label{fig:dataset}
\end{figure}

\section*{Results}
\subsection*{MedSAM: a foundation model for promptable medical image segmentation}
\begin{figure}[htbp]
\centering
\includegraphics[width=\linewidth]{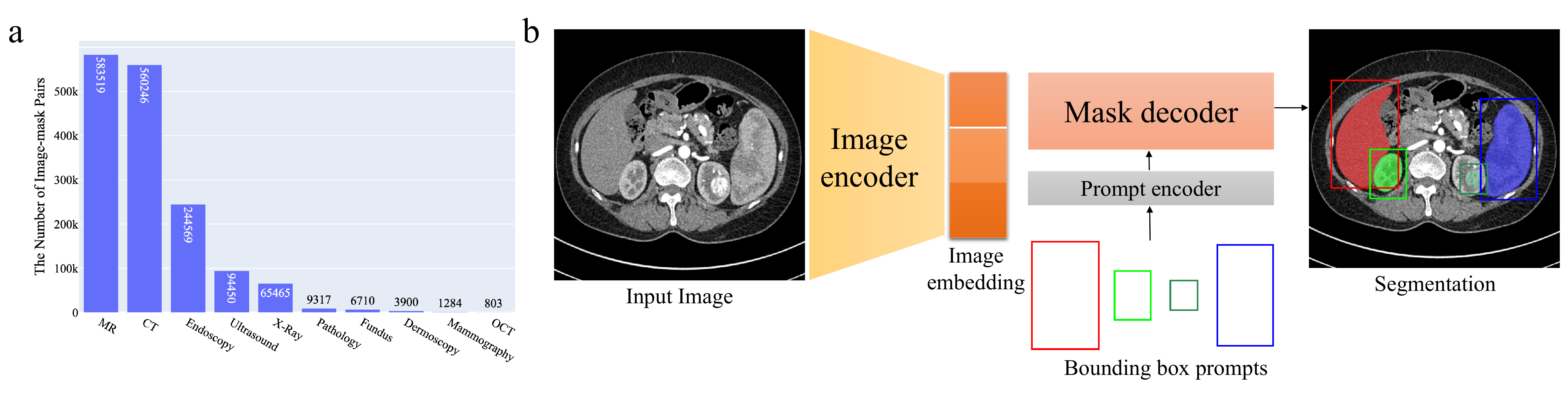}
\caption{\textbf{a,} The number of medical image-mask pairs in each modality. \textbf{b,} MedSAM is a promptable segmentation method where users can use bounding boxes to specify the segmentation targets. Source data are provided as a Source Data file.}
\label{fig:network}
\end{figure}

MedSAM aims to fulfill the role of a foundation model for universal medical image segmentation. A crucial aspect of constructing such a model is the capacity to accommodate a wide range of variations in imaging conditions, anatomical structures, and pathological conditions. 
To address this challenge, we curated a diverse and large-scale medical image segmentation dataset with 1,570,263 medical image-mask pairs, covering 10 imaging modalities, over 30 cancer types, and a multitude of imaging protocols (Fig.~\ref{fig:dataset}, Supplementary Table 1-4).  
This large-scale dataset allows MedSAM to learn a rich representation of medical images, capturing a broad spectrum of anatomies and lesions across different modalities.
Fig.~\ref{fig:network}a provides an overview of the distribution for images across different medical imaging modalities in the dataset, ranked by their total numbers. It is evident that Computed Tomography (CT), Magnetic Resonance Imaging (MRI), and endoscopy are the dominant modalities, reflecting their ubiquity in clinical practice. CT and MRI images provide detailed cross-sectional views of 3D body structures, making them indispensable for non-invasive diagnostic imaging. Endoscopy, albeit more invasive, enables direct visual inspection of organ interiors, proving invaluable for diagnosing gastrointestinal and urological conditions.
Despite the prevalence of these modalities, others such as ultrasound, pathology, fundus, dermoscopy, mammography, and Optical Coherence Tomography (OCT) also hold significant roles in clinical practice. The diversity of these modalities and their corresponding segmentation targets underscores the necessity for universal and effective segmentation models capable of handling the unique characteristics associated with each modality.

Another critical consideration is the selection of the appropriate segmentation prompt and network architecture. While the concept of fully automatic segmentation foundation models is enticing, it is fraught with challenges that make it impractical. One of the primary challenges is the variability inherent in segmentation tasks. For example, given a liver cancer CT image, the segmentation task can vary depending on the specific clinical scenario. One clinician might be interested in segmenting the liver tumor, while another might need to segment the entire liver and surrounding organs. Additionally, the variability in imaging modalities presents another challenge. Modalities such as CT and MR generate 3D images, whereas others like X-Ray and ultrasound yield 2D images.
These variabilities in task definition and imaging modalities complicate the design of a fully automatic model capable of accurately anticipating and addressing the diverse requirements of different users.

Considering these challenges, we argue that a more practical approach is to develop a promptable 2D segmentation model. The model can be easily adapted to specific tasks based on user-provided prompts, offering enhanced flexibility and adaptability. It is also able to handle both 2D and 3D images by processing 3D images as a series of 2D slices. Typical user prompts include points and bounding boxes and we show some segmentation examples with the different prompts in Supplementary Fig. 1. It can be found that bounding boxes provide a more unambiguous spatial context for the region of interest, enabling the algorithm to more precisely discern the target area. This stands in contrast to point-based prompts, which can introduce ambiguity, particularly when proximate structures resemble each other. Moreover, drawing a bounding box is efficient, especially in scenarios involving multi-object segmentation.  
We follow the network architecture in SAM~\cite{2023-SAM-Meta}, including an image encoder, a prompt encoder, and a mask decoder (Fig.~\ref{fig:network}b). The image encoder~\cite{ViT2020} maps the input image into a high-dimensional image embedding space. The prompt encoder transforms the user-drawn bounding boxes into feature representations via positional encoding~\cite{FourierPE-Nips20}. Finally, the mask decoder fuses the image embedding and prompt features using cross-attention~\cite{attention-Nips17} (Methods).

\begin{figure}[htbp]
\centering
\includegraphics[width=\linewidth]{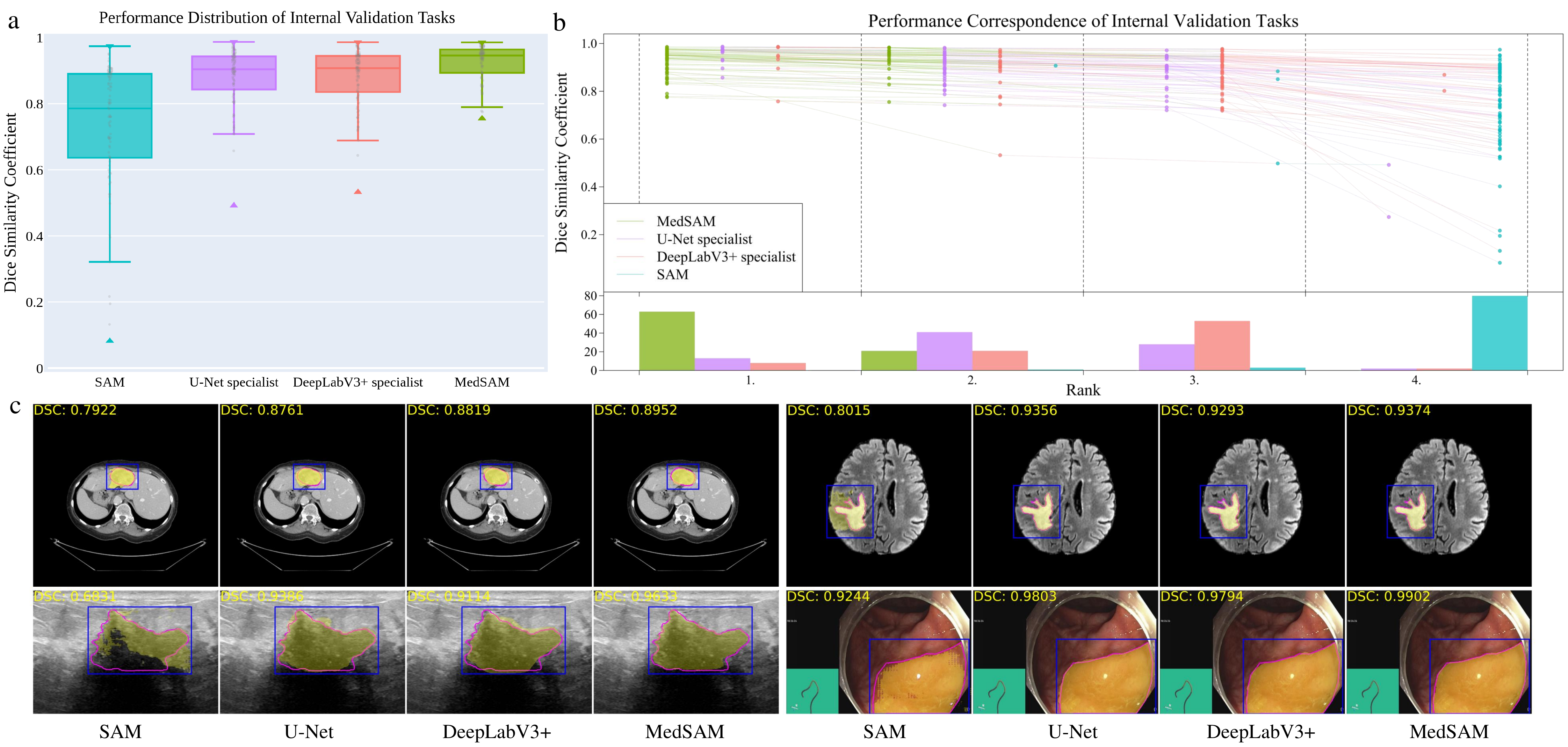}
\caption{\textbf{Quantitative and qualitative evaluation results on the internal validation set.} \textbf{a,} Performance distribution of 86 internal validation tasks in terms of median Dice Similarity Coefficient (DSC) score. The center line within the box represents the median value, with the bottom and top bounds of the box delineating the 25th and 75th percentiles, respectively. Whiskers are chosen to show the 1.5 of the interquartile range. Up-triangles denote the minima and down-triangles denote the maxima.
\textbf{b,} Podium plots for visualizing the performance correspondence of 86 internal validation tasks. Upper part: each colored dot denotes the median DSC achieved with the respective method on one task. Dots corresponding to identical tasks are connected by a line. Lower part: bar charts represent the frequency of achieved ranks for each method. MedSAM ranks in the first place on most tasks. 
\textbf{c,} Visualized segmentation examples on the internal validation set. The four examples are liver cancer, brain cancer, breast cancer, and polyp in Computed Tomography (CT), (Magnetic Resonance Imaging) MRI, ultrasound, and endoscopy images, respectively. Blue: bounding box prompts; Yellow: segmentation results. Magenta: expert annotations. 
Source data are provided as a Source Data file.
}
\label{fig:results}
\end{figure}

\subsection*{Quantitative and qualitative analysis}
We evaluated MedSAM through both internal validation and external validation. Specifically, we compared it to the SOTA segmentation foundation model SAM~\cite{2023-SAM-Meta} as well as modality-wise specialist U-Net~\cite{nnunet21} and DeepLabV3+~\cite{deeplabV3plus} models. Each specialized model was trained on images from the corresponding modality, resulting in 10 dedicated specialist models for each method. During inference, these specialist models were used to segment the images from corresponding modalities, while SAM and MedSAM were employed for segmenting images across all modalities (Methods).
The internal validation contained 86 segmentation tasks (Supplementary Table 5-8, Fig. 2), and Fig.~\ref{fig:results}a shows the median Dice Similarity Coefficient (DSC) score of these tasks for the four methods. 
Overall, SAM obtained the lowest performance on most segmentation tasks although it performed promisingly on some RGB image segmentation tasks, such as polyp (DSC: 91.3\%, interquartile range (IQR): 81.2-95.1\%) segmentation in endoscopy images. This could be attributed to SAM's training on a variety of RGB images, and the fact that many targets in these images are relatively straightforward to segment due to their distinct appearances.
The other three models outperformed SAM by a large margin and MedSAM has a narrower distribution of DSC scores of the 86 interval validation tasks than the two groups of specialist models, reflecting the robustness of MedSAM across different tasks. We further connected the DSC scores corresponding to the same task of the four models with the podium plot Fig.~\ref{fig:results}b, which is complementary to the box plot. In the upper part, each colored dot denotes the median DSC achieved with the respective method on one task. Dots corresponding to identical test cases are connected by a line. In the lower part, the frequency of achieved ranks for each method is presented with bar charts. It can be found that MedSAM ranked in first place on most tasks, surpassing the performance of the U-Net and DeepLabV3+ specialist models that have a high frequency of ranks with second and third places, respectively, In contrast, SAM ranked last place in almost all tasks. 
Fig.~\ref{fig:results}c (and Supplementary Fig. 9) visualizes some randomly selected segmentation examples where MedSAM obtained a median DSC score, including liver tumor in CT images, brain tumor in MR images, breast tumor in ultrasound images, and ployp in endoscopy images. SAM struggles with targets of weak boundaries, which is prone to under or over-segmentation errors. In contrast, MedSAM can accurately segment a wide range of targets across various imaging conditions, which achieves comparable of even better than the specialist U-Net and DeepLabV3+ models. 

\begin{figure}[htbp]
\centering
\includegraphics[width=\linewidth]{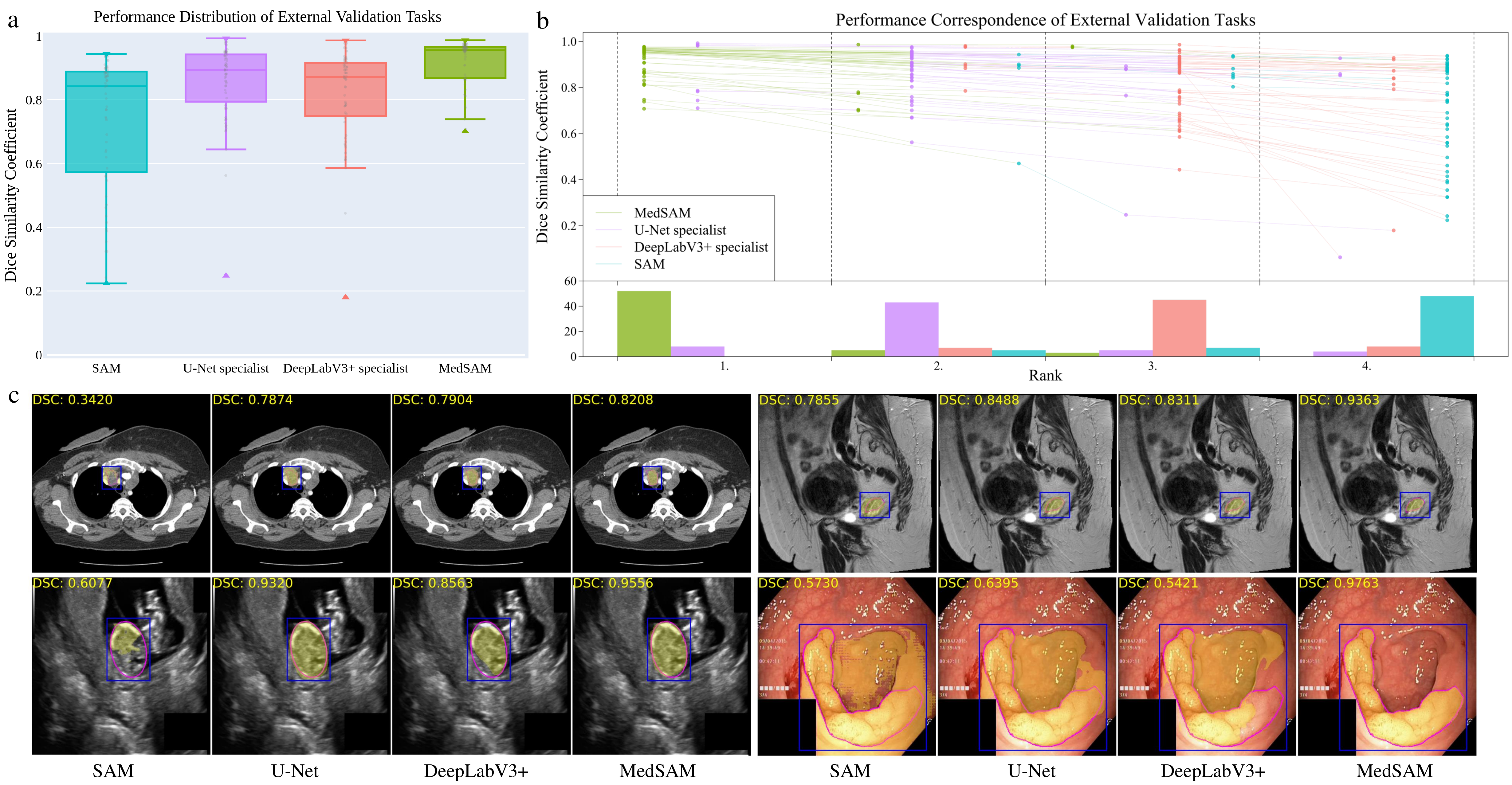}
\caption{\textbf{Quantitative and qualitative evaluation results on the external validation set.} \textbf{a,} Performance distribution of 60 external validation tasks in terms of median Dice Similarity Coefficient (DSC) score.
The center line within the box represents the median value, with the bottom and top bounds of the box delineating the 25th and 75th percentiles, respectively. Whiskers are chosen to show the 1.5 of the interquartile range. Up-triangles denote the minima and down-triangles denote the maxima. 
\textbf{b,} Podium plots for visualizing the performance correspondence of 60 external validation tasks. Upper part: each colored dot denotes the median DSC achieved with the respective method on one task. Dots corresponding to identical tasks are connected by a line. Lower part: bar charts represent the frequency of achieved ranks for each method. MedSAM ranks in the first place on most tasks. 
\textbf{c,} Visualized segmentation examples on the external validation set. The four examples are the lymph node, cervical cancer, fetal head, and polyp in CT, MR, ultrasound, and endoscopy images, respectively.
Source data are provided as a Source Data file.
}
\label{fig:results-external}
\end{figure}

The external validation included 60 segmentation tasks, all of which either were from new datasets or involved unseen segmentation targets (Supplementary Table 9-11, Fig. 10-12). Fig.~\ref{fig:results-external}a and b show the task-wise median DSC score distribution and their correspondence of the 60 tasks, respectively.
Although SAM continued exhibiting lower performance on most CT and MR segmentation tasks, the specialist models no longer consistently outperformed SAM (e.g., right kidney segmentation in MR T1-weighted images: 90.1\%, 85.3\%, 86.4\% for SAM, U-Net, and DeepLabV3+, respectively). This indicates the limited generalization ability of such specialist models on unseen targets. 
In contrast, MedSAM consistently delivers superior performance. For example, MedSAM obtained median DSC scores of 87.8\% (IQR: 85.0-91.4\%) on the nasopharynx cancer segmentation task, demonstrating 52.3\%, 15.5\%, and 22.7 improvements over SAM, the specialist U-Net, and DeepLabV3+, respectively. Significantly, MedSAM also achieved better performance in some unseen modalities (e.g., abdomen T1 Inphase and Outphase), surpassing SAM and the specialist models with improvements by up to 10\%.
Fig.~\ref{fig:results-external}c presents four randomly selected segmentation examples for qualitative evaluation, revealing that while all the methods have the ability to handle simple segmentation targets, MedSAM performs better at segmenting challenging targets with indistinguishable boundaries, such as cervical cancer in MR images (more examples are presented in Supplementary Fig. 13). 
Furthermore, we evaluated MedSAM on the multiple myeloma plasma cell dataset, which represents a distinct modality and task in contrast to all previously leveraged validation tasks. Although this task had never been seen during training, MedSAM still exhibited superior performance compared to the SAM (Supplementary Fig. 14), highlighting its remarkable generalization ability.

\begin{figure}[htbp]
\centering
\includegraphics[width=\linewidth]{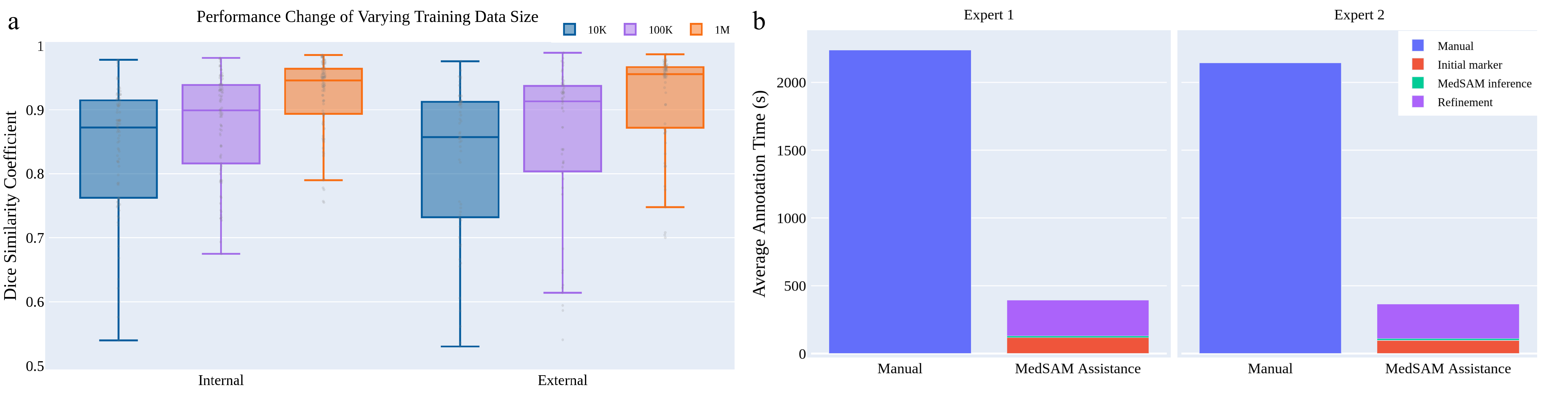}
\caption{\textbf{a,} Scaling up the training image size to one million can significantly improve the model performance on both internal and external validation sets. \textbf{b,} MedSAM can be used to substantially reduce the annotation time cost.
Source data are provided as a Source Data file.
}
\label{fig:clinical-app}
\end{figure}

\subsection*{The effect of training dataset size}
We also investigated the effect of varying dataset size on MedSAM’s performance because the training dataset size has been proven to be pivotal in model performance~\cite{heartNature23}. We additionally trained MedSAM on two different dataset sizes: 10,000 (10K) and 100,000 (100K) images and their performances were compared with the default MedSAM model. The 10K and 100K training images were uniformly sampled from the whole training set, to maintain data diversity. As shown in (Fig.~\ref{fig:clinical-app}a) (and Supplementary Table 12-14), the performance adhered to the scaling rule, where increasing the number of training images significantly improved the performance in both internal and external validation sets. 

\subsection*{MedSAM can improve the annotation efficiency}
Furthermore, we conducted a human annotation study to assess the time cost of two pipelines (Methods). For the first pipeline, two human experts manually annotate 3D adrenal tumors in a slice-by-slice way. For the second pipeline, the experts first drew the long and short tumor axes with the linear marker (initial marker) every 3-10 slices, which is a common practice in tumor response evaluation. Then, MedSAM was used to segment the tumors based on these sparse linear annotations. Finally, the expert manually revised the segmentation results until they were satisfied. We quantitatively compared the annotation time cost between the two pipelines (Fig.~\ref{fig:clinical-app}b).  The results demonstrate that with the assistance of MedSAM, the annotation time is substantially reduced by 82.37\% and 82.95\% for the two experts, respectively.

\section*{Discussion}
We introduce MedSAM, a deep learning-powered foundation model designed for the segmentation of a wide array of anatomical structures and lesions across diverse medical imaging modalities. MedSAM is trained on a meticulously assembled large-scale dataset comprised of over one million medical image-mask pairs. Its promptable configuration strikes an optimal balance between automation and customization, rendering MedSAM a versatile tool for universal medical image segmentation.

Through comprehensive evaluations encompassing both internal and external validation, MedSAM has demonstrated substantial capabilities in segmenting a diverse array of targets and robust generalization abilities to manage new data and tasks. Its performance not only significantly exceeds that of existing the state-of-the-art segmentation foundation model, but also rivals or even surpasses specialist models. 
By providing precise delineation of anatomical structures and pathological regions, MedSAM facilitates the computation of various quantitative measures that serve as biomarkers. For instance, in the field of oncology, MedSAM could play a crucial role in accelerating the 3D tumor annotation process, enabling subsequent calculations of tumor volume, which is a critical biomarker~\cite{RECIST09} for assessing disease progression and response to treatment. Additionally, MedSAM provides a successful paradigm for adapting natural image foundation models to new domains, which can be further extended to biological image segmentation~\cite{nm-seg-ma}, such as cell segmentation in light microscopy images~\cite{NeurIPS22CellSeg} and organelle segmentation in electron microscopy images~\cite{Ronald2023maester}.

While MedSAM boasts strong capabilities, it does present certain limitations. One such limitation is the modality imbalance in the training set, with CT, MRI, and endoscopy images dominating the dataset. This could potentially impact the model's performance on less-represented modalities, such as mammography. 
Another limitation is its difficulty in the segmentation of vessel-like branching structures because the bounding box prompt can be ambiguous in this setting. For example, arteries and veins share the same bounding box in eye fundus images. However, these limitations do not diminish MedSAM's utility. Since MedSAM has learned rich and representative medical image features from the large-scale training set, it can be fine-tuned to effectively segment new tasks from less-represented modalities or intricate structures like vessels.

In conclusion, this study highlights the feasibility of constructing a single foundation model capable of managing a multitude of segmentation tasks, thereby eliminating the need for task-specific models. MedSAM, as the inaugural foundation model in medical image segmentation, holds great potential to accelerate the advancement of new diagnostic and therapeutic tools, and ultimately contribute to improved patient care~\cite{AIradiomicsReview}.

\section*{Methods}

\subsection*{Dataset curation and pre-processing}
We curated a comprehensive dataset by collating images from publicly available medical image segmentation datasets, which were obtained from various sources across the internet, including the Cancer Imaging Archive (TCIA)~\cite{TCIA}, Kaggle, Grand-Challenge, Scientific Data, CodaLab, and segmentation challenges in the Medical Image Computing and Computer Assisted Intervention Society (MICCAI). All the datasets provided segmentation annotations by human experts, which have been widely used in existing literature (Supplementary Table 1-4). We incorporated these annotations directly for both model development and validation.

The original 3D datasets consisted of Computed Tomography (CT) and Magnetic Resonance (MR) images in DICOM, nrrd, or mhd formats. To ensure uniformity and compatibility with developing medical image deep learning models, we converted the images to the widely used NifTI format. Additionally, grayscale images (such as X-Ray and Ultrasound) as well as RGB images (including endoscopy, dermoscopy, fundus, and pathology images), were converted to the png format. Several exclusive criteria are applied to improve the dataset quality and consistency, including incomplete images and segmentation targets with branching structures, inaccurate annotations, and tiny volumes. 
Notably, image intensities varied significantly across different modalities. For instance, CT images had intensity values ranging from -2000 to 2000, while MR images exhibited a range of 0 to 3000. In endoscopy and ultrasound images, intensity values typically spanned from 0 to 255. To facilitate stable training, we performed intensity normalization across all images, ensuring they shared the same intensity range.

For CT images, we initially normalized the Hounsfield units using typical window width and level values.
The employed window width and level values for soft tissues, lung, and brain are (W:400, L:40), (W:1500, L:-160), and (W:80, L:40), respectively.
Subsequently, the intensity values were rescaled to the range of [0, 255].
For MR, X-Ray, ultrasound, mammography, and Optical Coherence Tomography (OCT)  images, we clipped the intensity values to the range between the 0.5th and 99.5th percentiles before rescaling them to the range of [0, 255]. Regarding RGB images (e.g., endoscopy, dermoscopy, fundus, and pathology images), if they were already within the expected intensity range of [0, 255], their intensities remained unchanged. However, if they fell outside this range, we utilized max-min normalization to rescale the intensity values to [0, 255].
Finally, to meet the model's input requirements, all images were resized to a uniform size of $1024\times1024\times3$. In the case of whole-slide pathology images, patches were extracted using a sliding window approach without overlaps. The patches located on boundaries were padded to this size with 0. 
As for 3D CT and MR images, each 2D slice was resized to $1024\times1024$, and the channel was repeated three times to maintain consistency.
The remaining 2D images were directly resized to $1024\times1024\times3$. Bi-cubic interpolation was used for resizing images, while nearest-neighbor interpolation was applied for resizing masks to preserve their precise boundaries and avoid introducing unwanted artifacts. These standardization procedures ensured uniformity and compatibility across all images and facilitated seamless integration into the subsequent stages of the model training and evaluation pipeline. 

\subsection*{Network architecture}
The network utilized in this study was built on transformer architecture~\cite{attention-Nips17}, which has demonstrated remarkable effectiveness in various domains such as natural language processing and image recognition tasks~\cite{ViT2020}. Specifically, the network incorporated a vision transformer (ViT)-based image encoder responsible for extracting image features, a prompt encoder for integrating user interactions (bounding boxes), and a mask decoder that generated segmentation results and confidence scores using the image embedding, prompt embedding, and output token.

To strike a balance between segmentation performance and computational efficiency, we employed the base ViT model as the image encoder since extensive evaluation indicated that larger ViT models, such as ViT Large and ViT Huge, offered only marginal improvements in accuracy~\cite{2023-SAM-Meta} while significantly increasing computational demands. Specifically, the base ViT model consists of 12 transformer layers~\cite{attention-Nips17}, with each block comprising a multi-head self-attention block and a Multilayer Perceptron (MLP) block incorporating layer normalization~\cite{layerNorm}. Pre-training was performed using masked auto-encoder modeling~\cite{MAE}, followed by fully supervised training on the SAM dataset~\cite{2023-SAM-Meta}. 
The input image ($1024\times1024\times3$) was reshaped into a sequence of flattened 2D patches with the size $16\times16\times3$, yielding a feature size in image embedding of $64\times64$ after passing through the image encoder, which is $16\times$ downscaled.
The prompt encoders mapped the corner point of the bounding box prompt to 256-dimensional vectorial embeddings~\cite{FourierPE-Nips20}. In particular, each bounding box was represented by an embedding pair of the top-left corner point and the bottom-right corner point. 
To facilitate real-time user interactions once the image embedding had been computed, a lightweight mask decoder architecture was employed. It consists of two transformer layers~\cite{attention-Nips17} for fusing the image embedding and prompt encoding, and two transposed convolutional layers to enhance the embedding resolution to $256\times256$. Subsequently, the embedding underwent sigmoid activation, followed by bi-linear interpolations to match the input size. 

\subsection*{Training protocol and experimental setting}
During data pre-processing, we obtained 1,570,263 medical image-mask pairs for model development and validation.
For internal validation, we randomly split the dataset into 80\%, 10\%, and 10\% as training, tuning, and validation, respectively. Specifically, for modalities where within-scan continuity exists, such as CT and MRI, and modalities where continuity exists between consecutive frames, we performed the data splitting at the 3D scan and the video level respectively, by which any potential data leak was prevented. For pathology images, recognizing the significance of slide-level cohesiveness, we first separated the whole-slide images into distinct slide-based sets. Then, each slide was divided into small patches with a fixed size of 1024x1024.
This setup allowed us to monitor the model's performance on the tuning set and adjust its parameters during training to prevent overfitting.
For the external validation, all datasets were hold-out and did not appear during model training. These datasets provide a stringent test of the model's generalization ability, as they represent new patients, imaging conditions, and potentially new segmentation tasks that the model has not encountered before. By evaluating the performance of MedSAM on these unseen datasets, we can gain a realistic understanding of how MedSAM is likely to perform in real-world clinical settings, where it will need to handle a wide range of variability and unpredictability in the data. The training and validation are independent.

The model was initialized with the pre-trained SAM model with the ViT-Base model. We fixed the prompt encoder since it can already encode the bounding box prompt. All the trainable parameters in the image encoder and mask decoder were updated during training. Specifically, the number of trainable parameters for the image encoder and mask decoder are 89,670,912 and 4,058,340, respectively. 
The bounding box prompt was simulated from the expert annotations with a random perturbation of 0-20 pixels. The loss function is the unweighted sum between Dice loss and cross-entropy loss, which has been proven to be robust in various segmentation tasks~\cite{nnunet21}. 
The network was optimized by AdamW~\cite{adamW} optimizer ($\beta_1$ = 0.9, $\beta_2$ = 0.999) with an initial learning rate of 1e-4 and a weight decay of 0.01. The global batch size was 160 and data augmentation was not used. 
The model was trained on 20 A100 (80G) GPUs with 150 epochs and the last checkpoint was selected as the final model.

Furthermore, to thoroughly evaluate the performance of MedSAM, we conducted comparative analyses against both the state-of-the-art segmentation foundation model SAM~\cite{2023-SAM-Meta} and specialist models (i.e., U-Net~\cite{nnunet21} and DeepLabV3+~\cite{deeplabV3plus}). The training images contained 10 modalities: CT, MR, chest X-Ray (CXR), dermoscopy, endoscopy, ultrasound, mammography, OCT, and pathology, and we trained the U-Net and DeepLabV3+ specialist models for each modality. There were 20 specialist models in total and the number of corresponding training images was presented in Supplementary Table 5. 
We employed the nnU-Net to conduct all U-Net experiments, which can automatically configure the network architecture based on the dataset properties. In order to incorporate the bounding box prompt into the model, we transformed the bounding box into a binary mask and concatenated it with the image as the model input. This function was originally supported by nnU-Net in the cascaded pipeline, which has demonstrated increased performance in many segmentation tasks by using the binary mask as an additional channel to specify the target location. The training settings followed the default configurations of 2D nnU-Net. Each model was trained on one A100 GPU with 1000 epochs and the last checkpoint was used as the final model.
The DeepLabV3+ specialist models used ResNet50~\cite{resnet16} as the encoder. Similar to ~\cite{heart-nature}, the input images were resized to $224 \times 224 \times 3$. The bounding box was transformed into a binary mask as an additional input channel to provide the object location prompt. Segmentation Models Pytorch (0.3.3)~\cite{Iakubovskii:2019} was used to perform training and inference for all the modality-wise specialist DeepLabV3$+$ models. Each modality-wise model was trained on one A100 GPU with 500 epochs and the last checkpoint was used as the final model. 
During the inference phase, SAM and MedSAM were used to perform segmentation across all modalities with a single model. In contrast, the U-Net and DeepLabV3+ specialist models were used to individually segment the respective corresponding modalities.

A task-specific segmentation model might outperform a modality-based one for certain applications. Since U-Net obtained better performance than DeepLabV3+ on most tasks, we further conducted a comparison study by training task-specific U-Net models on four representative tasks, including liver cancer segmentation in CT scans, abdominal organ segmentation in MR scans, nerve cancer segmentation in ultrasound, and polyp segmentation in endoscopy images. 
The experiments included both internal validation and external validation. For internal validation, we adhered to the default data splits, using them to train the task-specific U-Net models and then evaluate their performance on the corresponding validation set. For external validation, the trained U-Net models were evaluated on new datasets from the same modality or segmentation targets. In all these experiments, MedSAM was directly applied to the validation sets without additional fine-tuning. 
As shown in Supplementary Fig. 15, while task-specific U-Net models often achieved great results on internal validation sets, their performance diminished significantly for external sets. In contrast, MedSAM maintained consistent performance across both internal and external validation sets. This underscores MedSAM's superior generalization ability, making it a versatile tool in a variety of medical image segmentation tasks. 

\subsection*{Loss function}
We used the unweighted sum between cross-entropy loss and Dice loss~\cite{vnet-diceloss} as the final loss function since it has been proven to be robust across different medical image segmentation tasks~\cite{SegLossOdyssey}.
Specifically, let $S, G$ denote the segmentation result and ground truth, respectively. $s_i, g_i$ denote the predicted segmentation and ground truth of voxel $i$, respectively. $N$ is the number of voxels in the image $I$.
Binary cross-entropy loss is defined by
\begin{equation}
    L_{BCE} = -\frac{1}{N}\sum_{i=1}^{N}
    \left[
    g_{i} \log s_{i} +
    (1 - g_{i}) \log (1 - s_{i})
    \right]
    ,
\end{equation}
and dice loss is defined by 
\begin{equation}\label{eq:DiceV1}
    L_{Dice} = 1- \frac{2\sum_{i=1}^{N}g_{i}s_{i}}{\sum_{i=1}^{N}(g_{i})^2 + \sum_{i=1}^{N}(s_i)^2}.
\end{equation}
The final loss $L$ is defined by
\begin{equation}
    L = L_{BCE} + L_{Dice}.
\end{equation}

\subsection*{Human annotation study}
The objective of the human annotation study was to quantitatively evaluate how MedSAM can reduce the annotation time cost. Specifically, we used the recent adrenocortical carcinoma CT dataset~\cite{ahmed2020radiomic, adrental2, TCIA}, where the segmentation target, adrenal tumor, was neither part of the training nor of the existing validation sets. 
We randomly sampled 10 cases, comprising a total of 733 tumor slices requiring annotations. Two human experts participated in this study, both of whom are experienced radiologists with 8 and 6 years of clinical practice in abdominal diseases, respectively. Each expert generated two groups of annotations, one with the assistance of MedSAM and one without.

In the first group, the experts manually annotated the 3D adrenal tumor in a slice-by-slice manner. Annotations by the two experts were conducted independently, with no collaborative discussions, and the time taken for each case was recorded.  
In the second group, annotations were generated after one week of cooling period. The experts independently drew the long and short tumor axes as initial markers, which is a common practice in tumor response evaluation. This process was executed every 3-10 slices from the top slice to the bottom slice of the tumor. Then, we applied MedSAM to segment the tumors based on these sparse linear annotations, including three steps.
\begin{itemize}
    \item Step 1. For each annotated slice, a rectangle binary mask was generated based on the linear label that can completely cover the linear label.
    \item Step 2. For the unlabeled slices, the rectangle binary masks were created through interpolation of the surrounding labeled slices.
    \item Step 3. We transformed the binary masks into bounding boxes and then fed them along with the images into MedSAM to generate segmentation results. 
\end{itemize}
All these steps were conducted in an automatic way and the model running time was recorded for each case. 
Finally, human experts manually refined the segmentation results until they met their satisfaction. 
To summarize, the time cost of the second group of annotations contained three parts: initial markers, MedSAM inference, and refinement. 
All the manual annotation processes were based on ITK-SNAP~\cite{ITK-SNAP}, an open-source software designed for medical image visualization and annotation.

\subsection*{Evaluation metrics}
We followed the recommendations in Metrics Reloaded~\cite{metric-reload} and used Dice Similarity Coefficient and Normalized Surface Distance (NSD) to quantitatively evaluate the segmentation results. 
DSC is a region-based segmentation metric, aiming to evaluate the region overlap between expert annotation masks and segmentation results, which is defined by 
\begin{equation*}
    DSC(G, S) = \frac{2|G\cap S|}{|G| + |S|},
\end{equation*}
NSD~\cite{NSD} is a boundary-based metric, aiming to evaluate the boundary consensus between expert annotation masks and segmentation results at a given tolerance, which is defined by
\begin{equation*}
    NSD(G, S) = \frac{|\partial G\cap B_{\partial S}^{(\tau)}| + |\partial S\cap B_{\partial G}^{(\tau)}|}{|\partial G| + |\partial S|},
\end{equation*}
where  $B_{\partial G}^{(\tau)} = \{x\in R^3 \, | \, \exists \tilde{x}\in \partial G,\, ||x-\tilde{x}||\leq \tau \}$, $B_{\partial S}^{(\tau)} = \{x\in R^3 \,|\, \exists \tilde{x}\in \partial S,\, ||x-\tilde{x}||\leq \tau \}$  denote the border region of the expert annotation mask and the segmentation surface at tolerance $\tau$, respectively. In this paper, we set the tolerance $\tau$ as 2.

\subsection*{Statistical analysis}
To statistically analyze and compare the performance of the aforementioned four methods (MedSAM, SAM, U-Net, and DeepLabV3+ specialist models), we employed the Wilcoxon signed-rank test. This non-parametric test is well-suited for comparing paired samples and is particularly useful when the data does not meet the assumptions of normal distribution. This analysis allowed us to determine if any method demonstrated statistically superior segmentation performance compared to the others, providing valuable insights into the comparative effectiveness of the evaluated methods. The Wilcoxon signed-rank test results are marked on the DSC and NSD score tables (Supplementary Table 6-11). 

\subsection*{Software Utilized}
All code was implemented in Python (3.10) using Pytorch (2.0) as the base deep learning framework. We also used several python packages for data analysis and results visualization, including connected-components-3d (3.10.3), SimpleITK (2.2.1), nibabel (5.1.0), torchvision (0.15.2), numpy (1.24.3), scikit-image (0.20.0), scipy (1.10.1), and pandas (2.0.2), matplotlib (3.7.1), opencv-python (4.8.0), ChallengeR (1.0.5), and plotly (5.15.0). Biorender was used to create Fig. 1. 

\subsection*{Data availability}
The training and validating datasets used in this study are available in the public domain and can be downloaded via the links provided in the Supplementary Table 16-17. Source data are provided with this paper in the Source Data file. We confirmed that All the image datasets in this study are publicly accessible and permitted for research purposes.

\subsection*{Code availability}
The training script, inference script, and trained model have been publicly available at \url{https://github.com/bowang-lab/MedSAM}. A permanent version is released on Zenodo~\cite{medsam-zenodo}. 
\backmatter

\bmhead{Acknowledgments}
This work was supported by the Natural Sciences and Engineering Research Council of Canada (NSERC, RGPIN-2020-06189 and DGECR-2020-00294) and CIFAR AI Chair programs.
The authors of this paper highly appreciate all the data owners for providing public medical images to the community. We also thank Meta AI for making the source code of segment anything publicly available to the community. 
This research was enabled in part by computing resources provided by the Digital Research Alliance of Canada. 

\subsection*{Author Contributions}
Conceived and designed the experiments: J.M. Y.H., C.Y., B.W.
Performed the experiments: J.M. Y.H., F.L., L.H., C.Y. 
Analyzed the data: J.M. Y.H., F.L., L.H., C.Y., B.W. 
Wrote the paper: J.M. Y.H., F.L., L.H., C.Y., B.W.
All authors have read and agreed to the published version of the manuscript.

\subsection*{Competing Interests}

The authors declare no competing interests


\bibliography{sn-bibliography}


\end{document}